\documentstyle[aps,floats,graphicx]{revtex}

\begin{document}

\newcommand{\bec}{\begin{center}}
\newcommand{\ec}{\end{center}}
\newcommand{\be}{\begin{equation}}
\newcommand{\ee}{\end{equation}}
\newcommand{\beqn}{\begin{eqnarray}}
\newcommand{\eeqn}{\end{eqnarray}}
\newcommand{\bet}{\begin{table}}
\newcommand{\ent}{\end{table}}
\newcommand{\bib}{\bibitem}

\wideabs{

\title{
Anomalous inter-layer atomic transport and
the low-temperature amplification of surface instability in Al(111) 
%Low-temperature amplification of surface instability and anomalous interdiffusion 
%Surface instability induced superdiffusion with oscillatory interaction: Pt on Al(111)
%Superdiffusion with an attractive surface induced oscillatory interaction: Pt on Al(111)
%°Superdiffusion with a surface vortex matter induced oscillatory interaction: Pt on Al(111)
%Anomalous diffusion during thin film growth
%Size-mismatch induced enhancement of interdiffusion 
}

\author{P. S\"ule} 
  \address{Research Institute for Technical Physics and Material Science,\\
Konkoly Thege u. 29-33, Budapest, Hungary,sule@mfa.kfki.hu,www.mfa.kfki.hu/$\sim$sule,\\
}
%\email{sule@mfa.kfki.hu}

\date{\today}

\maketitle

\begin{abstract}
Anomalous inter-layer atomic transport of deposited impurity atoms on Al(111) has been found
by constant temperature molecular dynamics simulations.
The low-energy deposition of Pt on Al(111) leads to oscillatory adsorbate-substrate interaction
and to a low-temperature ballistic injection of the deposited particles to below the topmost layer.
The ultrafast injection of a Pt atom coincides with the ejection of
a substrate atom to the surface (ballistic replacement or exchange mechanism). This is in agreement with the experimental
findings in which thin film rich in Al has been found on the surface after deposition of
few MLs of Pt.
We attribute the anomalous inter-layer transport to the size-mismatched impurity/host
interaction and we point out the role of atomic size mismatch
in biasing towards intermixing.
The deposition induced low-temperature disordering of surface Al atoms with few transient Al adatoms has also been observed.
The ultrafast injection of the impurity particles to the substrate is assisted by the transient out-of-plane 
and lateral circulating motion of few surface atoms arranged nearly in a hexagonal symmetry.
The atomic injection of Pt can be regarded as a superexchange process driven by the
transient and collective motion of few surface atoms.
A chaotic surface state assisted mechanism could also be a possible explanation
of the unexpected phenomenon.

\\

%}
\end{abstract}
}

\section{Introduction}

 The understanding of the fundamental atomistic transport processes leading to the formation and morphological evolution of nanoscale surface features is lacking \cite{Schukin}.
% The understanding of the fundamental phenomena leading to the formation, stability, and morphological evolution of nanoscale features is lacking \cite{Schukin}.
Many classical macroscopic (continuum and mesoscopic) models for diffusion and morphological evolution lose their validity 
in the nanoscale \cite{Schukin,Beke}.
%As the dimensions of the surface features is reduced to the nanoscale, many classical macroscopic (continuum and mesoscopic) models for diffusion and morphological evolution lose their validity \cite{Schukin,Beke}.
Therefore the atomistic level understanding of the driving forces and mechanisms governing atomic transport involved in the synthesis and organisation of nanoscale features in solid state materials
including atomic intermixing,
is inevitable \cite{Schukin}.

  The most well known atomic diffusion mechanisms of adatoms on solid surfaces are the
hopping diffusion on the surface from one site to another over a bridge site and the atomic site exchange when an adatom
enters the topmost layer and a surface atom becomes an adatom \cite{Michely}.
   The atoms deposited at surfaces could undergo, however, few exotic (or anomalous) transport processes
such as random walk or Levy flight \cite{Michely,Anomaldiff,Levy}, Friedel-type adsorbate-adsorbate oscillatory interaction
driven relaxations \cite{Michely,Luo,Meyerheim,Polop,oscill}, intermixing and surface
alloying between immiscible elements \cite{Meyerheim,Tersoff,Stepanyuk2,Levanov}, 
long range surface mediated lateral mass transport within the topmost layer \cite{Bulou}
and very fast kinetics with low diffusion barrier driven by quantum size effects \cite{Chan}.
The low-energy ion-bombardment induced ejection of Al atoms to the Al(111) surface
results in adatoms whith ultrafast motion on the surface leading to the
formation of nanodots \cite{Sule_SUCI}.
These atomic migrations can be characterized by the lateral transport of atoms
on the surface or whitin the topmost layer. Inter-layer transport occurs only
between the surface and the topmost layer or at step edges, however, the most of these processes
can not usually be considered as anomalous, e.g. the rate of the process
can be described within the framework of the Arrhenius equation.
Anomalous atomic processes are athermal and the mean free path of the atomic 
jumps can reach the few times of the nearest neighbor distance \cite{Michely,Anomaldiff}.
Also, the mean square of atomic displacements scales nonlinearily in the course of time
of diffusion \cite{Anomaldiff,Sule_submitted} or the nanoscale growth rate of interfaces follows linear growth kinetics (nonparabolic growth) 
\cite{Beke}.

 However, only few examples are known for inter-layer mass transport (atomic jumps normal to the surface) which can be considered
as anomalous. 
Even in these articles the authors not always realized the anomalous nature
of inter-layer transport.
 The assistance of instantenously high kinetic energy to inter-layer transport
during the deposition of Au on Ag surface has been reported \cite{Fichthorn}.
Interdiffusive Stranski-Krastanov growth mode in
an analysis of scanning tunneling microscopy data 
and the burrowing of Au particles in Ag(110) when annealing has been applied during simulations (replacement diffusion mechanism) have also been found \cite{Haftel1}.
In another study a complicated exchange mechanism is shown with ballisticaly
injected deposited Au on Ag(111) \cite{Haftel2}.
Sprague et al. has been published interface mixing on impact
(athermal mixing) for the case of Pt deposition on Cu and which is explained by the attractive potentials of
the substrate atoms and a large thermodynamic bias toward film-substrate mixing \cite{Sprague}.
%Large capillary forces driven burrowing of Co clusters on the Cu(001) surface has also been found
%recently \cite{Stepanyuk2,Levanov}.
The mechanism of the seemingly different transient atomic migrations could hopefully be explained in a common framework.

% The abovementioned and many other experimental and theroretical results can not be explained as
%conventional diffusion. 
  The understanding of how adsorbed impurities and growing film constituents
affect interdiffusion to the substrate needs further studies of the
atomistic mechanism of intermixing.
For instance, using vapor deposition of various transition metals on Al, it has been
found that the intermixing (IM) length is anomalously large in certain cases \cite{Buchanan}.
It has also been characterized that IM is not driven by bulk diffusion parameters \cite{Buchanan} nor by
bulk thermochemistry (such as heats of alloying) during ion-beam mixing \cite{Sule_PRB05,Sule_NIMB04}.
The segregation of Al during vapor deposition of Pt on polycrystalline Al together with the fast reactive diffusion of Pt to the bulk have also
been found \cite{Buchanan,Barna}.
In another study
the adsorption of Pt on polycrystalline Al leads to the formation of surface alloys \cite{Rodriguez}
which are rich in Al 
\cite{Barna2}.

  Anharmonic effects, surface disordering and premelting has also been studied intensively
at fcc (111) metal surfaces \cite{premelting,Rahman}.
It has also been concluded that surface thermal expansion and lattice dynamics
might be anomalous of the (111) surface layer of Al \cite{premelting,Forsblom}
hence Al can be described by unusual surface instability. It would be
interesting if e.g. the surface of Al would be subjected to small external perturbation
such as the deposition of an impurity atom.
Owing to the anharmonic surface behavior of Al(111)
the amplification of surface instability could be expected.

 A number of unconventional atomic transport phenomena have been explained by atomic size-mismatch (ASM) 
of the constituents \cite{Tersoff,Sule_submitted,Okamoto,Jalali,Dalmas,Li,Sheng,Qi,Park}.
Surface alloying and intermixing between many immiscible and even miscible elements 
can not solely be explained by thermodynamic and chemical forces such as heat of alloying and mixing or cohesive energies
 \cite{Tersoff,Stepanyuk2,Levanov,Sule_PRB05,Sule_NIMB04}.
The puzzling nature of intermixing (IM) might be due to still an unknown and an unconventional mechanism
which has not been explored yet or overlooked in the last decades.
In our previous reports we explained IM as an interfacial anisotropy driven process in diffusion couples
and pointed out the role of mass and size-anisotropy during
ion-bombardment induced interdiffusion \cite{Sule_SUCI,Sule_submitted,Sule_PRB05}.

  In the present article, we show that anomalous atomic transport occurs 
during vapor deposition in size-mismatched systems, such as during
 the thin film growth of Pt and other transition metals on Al(111). 
The interdiffusion of the deposited particle is driven by an oscillatory interaction
with the surface atoms of the substrate.
Moreover, we point out that the anomalous character of the atomic transport
of the deposit can be tuned by adjusting the ASM of the constituents.
We present an exchange mechanism with a
concerted motion of few surface atoms 
for entering the uppermost layer by the deposited particle.
Also, we show that large atomic vibrational displacements and anharmonic effects
might occur not only at high-temperatures close to the melting point, but also at
ultra-low temperatures induced by impurities at the surface.

\section{The simulation method}

\subsection{General properties}

 We give the detailed outline of the employed simulation approach. 
 Classical constant volume tight-binding molecular dynamics simulations \cite{CR} were used to simulate 
soft landing and vapor deposition of Pt atoms on Al(111) and also on other substrates (such as Cu(111)) surface at $\sim 0$ K
using the PARCAS code \cite{Nordlund_ref}.
The PARCAS MD code has widely been used for the study of various atomic transport phenomena
in the last few years \cite{Sule_PRB05,PARCAS}.
A variable timestep
and the Berendsen temperature control is used \cite{Allen,Frenkel}. 
The simulation uses the Gear's predictor-corrector algorithm to calculate
atomic trajectories \cite{Allen}.
The maximum time step of $0.05$ fs is used.
We consider the coupling of our simulation cell to a heat bath by inserting stochastic and
friction terms to the equation of motion yielding a Langevin type equation (see details
in refs. \cite{Allen,Frenkel,Berendsen}).
The equation of motion can be written as follows,
\be
 m_i \dot{v_i}= {\bf F_i} - m_i \gamma_i v_i + R(t),
\ee
where $\bf{F_i}$ is the force on atom $i$ and 
$R(t)$ is a Gaussian stochastic variable.
The damping constants $\gamma_i$ determine the strength of coupling to the thermostat.
The global coupling to the heat bath can be adjusted by the so called
Berendsen temperature which we set to $70$ K.
The system couples to a heat bath not only globally via the damping constant but is also locally
subjected to random noise.
Stochastic forces and damping applied at the cell borders (lateral $x-y$ boundaries as heat sink) of the simulation cell
to maintain constant temperature conditions and the thermal equilibrium of the entire
system (coupling to the heat bath). The top of the simulation cell is left free (the free surface) for
the deposition of Pt atoms.
The bottom layers
are held fixed in order to avoid the rotation of the cell. 
system.
  For simulating deposition it is appropriate to use temperature control
at the cell borders.
This is because it is physically correct that potential energy becomes
kinetic energy on impact, i.e. heats the lattice. This heating should
be allowed to dissipate naturally, which means temperature control should not
be used at the impact point. 
%We also employ temperature control on all the atoms (not only at the borders).
Periodic boundary conditions are imposed laterarily.
The observed anomalous transport processes are also observed without
periodic boundary conditions and temperature control.
Further details are given in \cite{Nordlund_ref} and details specific to the current system in recent
communications \cite{Sule_PRB05,Sule_NIMB04}.

  The size of the simulation cell is $80 \times 80 \times 42$ $\hbox{\AA}^3$ including
 $16128$ atoms (with a fcc lattice).
$15$ active MLs are supported on $3$ fixed bottom monolayers (MLs).
We find no dependence of the anomalous atomic transport properties of the deposited atoms on the
finite size of the simulation cell.
Deposited atoms were initialized normal to the (111) surface with randomly
selected lateral positions $4-5$ $\hbox{\AA}$ above the surface.

 The kinetic energy of the deposited particles are
$\sim 0.1$ eV or in the case of ultrasoft landing it is nearly zero eV.
The impurity (deposited) particle is placed $4-5$ $\hbox{\AA}$ above the (111) surface of the substrate.
No acceleration of the deposited particle is observed above $6$ $\hbox{\AA}$ the
surface.
We also analyze the acceleration of the deposited particles and calculate
the arrival energy at the surface of the substrate.
Using the history (movie) file we collect the atomic positions of the moving substrate atoms
which have kinetic energy above a certain threshold value in order to get
the pattern of atomic trajectories during the deposition events.
At $0$ K this
value is $\sim 0.01$ eV.
In order to make a statistics of impact events we generated $100$ events with randomly
varied impact positions.

  In order to follow the time evolution of atomic motions the
mean square of the atomic displacements $\langle R^2 \rangle$ (MSD) has been calculated.
$\langle R^2 \rangle = 
\sum_i^N [{\bf r_i}(t)-{\bf r_i}(t=0)]^2$, as obtained by molecular dynamics simulations and (${\bf r_i}(t)$ is the position vector of atom 'i' at time $t$) and $N$ is the number of atoms included in the sum, respectively.
We also calculate $\langle R^2 \rangle$ for the impurity atom only, hence in this case
$\langle R^2 \rangle = ({\bf r_i}(t)-{\bf r_i}(t=0))^2$.
In other cases lateral or vertical (out-of-plane) components are included in
$\langle R^2 \rangle$ for the substrate atoms.
We folow the evolution of $\langle R^2 \rangle$ during few events in order to get a 
reasonable statistics of atomic motions.

\subsection{The interaction potentials}

 We use the many-body Cleri-Rosato (CR) tight-binding second-moment approximation (TB-SMA) interaction potential to describe interatomic interactions \cite{CR}.
The CR potential is formally analogous to the embedded atomic method (EAM, \cite{EAM}) formalism, e.g. 
the potential energy of an atom is given as a sum of repulsive pair potentials for the
neighboring atoms (usually for the first or second neighbors and a cutoff is imposed out of this
region) and an embedding energy that is a function of the local electron density
given as follows \cite{EAM},
\be
E_{tot}= \frac{1}{2}\sum_{ij} V(r_{ij}) +\sum_i F[\rho_{i}],
\label{eq1}
\ee
where $r_{ij}$ is the distance between atoms $i$ and $j$ and its neighbors.
There are many functional forms are available for the density $\rho_{i}$ and for the embedding
function $F[\rho_{i}]$ \cite{EAM}.
%In the code PARCAS \cite{Nordlund_ref} the forces have been calculated
%using a built-in functional derivative of Eq. (1).
We utilize EAM functional forms in the code for $F[\rho_{i}]$ and for the density $\rho$
similar to that given in refs. \cite{EAM,Haftel}.
The EAM routine in the code employs a cubic spline interpolation for the evaluation
of the EAM potentials and their derivatives (forces) starting from various kind of input potentials
given in discrete points as a function of $r_{ij}$ (the number of points per functions
is $5000$ in this study).

%The potential is empirical and is obtained by fitting bulk material properties of lattice
%parameter, cohesive energy, elastic constants, sublimation energy, vacancy formation energy.
 Within the second moment tight-binding approach, the band energy
(the attractive part of the potential) reads,
\be
 E_b^i=-\biggm[ \sum_{j, r_{ij} < r_c} \xi^2 exp \biggm[-2q \biggm(\frac{r_{ij}}{r_0}
-1 \biggm) \biggm] \biggm]^{1/2},
\ee
where $r_c$ is the cutoff radius of the interaction and $r_0$ is the first neighbor distance (atomic size parameter).
%Using $\Theta$ we impose cutoff in the embedding function.
%$\Theta=0$, if $r_{ij} > r_c$ and $\Theta=1$, if $r_{ij} \le r_c$.
%\beqn
%v_s([\rho],{\bf r}) = v_s([\rho,\{u_i\},\{\epsilon_i\}],{\bf r}) && \nonumber \\
%= v_{ext}([\rho],{\bf r}) + v_H([\rho],{\bf r}) && \nonumber \\ +v_{xc}([\rho,\{u_i\},\{\epsilon_i\}],{\bf r})
%\eeqn

The repulsive term is a Born-Mayer type phenomenological core-repulsion term:
\be
E_r^i=A \sum_{j, r_{ij}<r_c} exp \biggm[-p \biggm(\frac{r_{ij}}{r_0}-1 \biggm) \biggm].
\ee
The parameters ($\xi, q, A, p, r_0$) are fitted to experimental values of the cohesive energy,
the lattice parameter, the bulk modulus and the elastic constants $c_{11}$, $c_{12}$ and $c_{44}$ \cite{CR}.
The summation over $j$ is extended up to fifth neighbors for fcc structures \cite{CR}.
The total cohesive energy of the system is
\beqn
E_c=-\sum_{i,j,r_{ij}<r_c, i \ne j} \int_0^{\infty} \frac{\partial U(\bf{r})}{\partial \bf{r}} \biggm|_{r=r_{ij}} \frac{\overline{r_{ij}}}{r_{ij}} d \bf{r} && \nonumber \\
 = \sum_i (E_r^i+E_b^i),~~~~~~~~~~~~~~~~~~~~~~~~~~~~~~
\eeqn
where $U(\bf{r})$ is the total potential energy of the entire system as a function of the
space coordinate (radius vector) $\bf{r}$ since the system is energy conservative, the space integral
over the Newtonian interatomic forces gives the total energy of the simulation cell.
$\bf{r}$ can be replaced by the internuclear separation $r_{ij}$ in the pair interaction term
$V(r_{ij})$ and by the position vector of the electron density in the density function
$\rho_i(r)$ in Eq. (2).
Formally we give all the terms in Eqs. (3)-(4) in the same way as for
EAM potentials except that 
$F[\rho_{i}]$ is calculated in a tight-binding form as given in Eq. (3). We prefer to use the CR potential
because $r_0$ can be tuned in this case which is proportional to the atomic size-mismatch
in the heteronuclear potential (the proportionality will be discussed later in section III.). 
Using the CR potential we consider the interaction between two atoms and the interaction
with their local environment usually up to the second neighbors.
Out of this region a potential cutoff is imposed.
The cutoff radius $r_c$ is taken as the second neighbor distance for all the interactions.
We tested the Al-Al and the Al-Pt potential at cutoff radius with third and larger neighbor distances and
found no considerable change in the results.
The CR potential gives the physically reasonable representation of metals and computationally is efficient
hence nanoscale atomic clusters with large number of atoms (up to millions of atoms) can be simulated.
This type of a potential gives a very good description of lattice vacancies, including migration
properties and a reasonable description of solid surfaces and melting \cite{CR}.
Since the present work is mostly associated with the elastic properties,
melting behaviors, surface, interface and migration energies, we believe the model used should be suitable for this study.
In a recent report we found that the CR potential describes properly Al
and Al/Pt \cite{Sule_SUCI} and provides reasonable results for the ion-bombardment of
Al(111) in agreement with the experimental results \cite{Michely}.
The CR potential correctly provides the adatom binding and dimerization energies, adatom island
formation upon ion-bombardment in agreement with experiment \cite{Busse_MD}.
Recently it has also been shown, that the CR potential remarkably well describes
diffusion in liquid Al \cite{Li2,Alemany} and energetic deposition of
Al clusters on Al \cite{Kang}.
In order to be more convincing in the accuracy of the TB-SMA model we also used another
parameterization set for Eqs. (3)-(4) obtained by first principles augmented-plane-wave
calculations \cite{Papanicolau}.
This parameterization offers a satisfactory agreement with the experimental data
for thermal expansion coefficient, the temperature dependence of the
atomic mean-square of displacement and the phonon density of states of compounds \cite{Papanicolau}.
However, we find no serious difference in the final results obtained for
deposition between the CR and the parameterization of Papanicolau {\em et al.}.
Therefore we use the original CR parameterization set for Al.

Despite the empirical nature of the EAM and CR potentials their accuracy seems to be
sufficient for studying materials under nonequilibrium conditions such as
anharmonic effects and thermal expansion of various fcc metals \cite{premelting,Rahman}.
In the present paper we use these potentials for the investigation of
low-temperature disordering processes with transient surface atomic vibrations for which
the EAM and CR potentials should be adequate \cite{premelting,Sule_SUCI}.

  For the crosspotential of substrate atoms and Pt we employ an interpolation scheme \cite{Sule_SUCI,Sule_PRB05,geometric}
using 
the geometrical mean of the elemental energy constants and the harmonic mean for the screening
length of Eqs. (3)-(4) are taken as in refs. \cite{geometric,AB}.
The CR elemental potentials and the interpolation scheme for heteronuclear interactions
have widely been used for MD simulations \cite{Stepanyuk2,Levanov,Sule_PRB05,Dalmas,Goyhenex}.
Recently CR interpolated crosspotential has also successfully been used for Ti/Pt in agreement with
our experimental results \cite{Sule_submitted}.
The scaling factor $r_0$ (the heteronuclear first neighbor distance) is calculated as the average of the elemental first neighbor distances. 
The AlPt potential is fitted to the
measured effective heat of mixing in the cubic AlPt ($\Delta H \approx -100$ kJ/mol) with a
melting temperature of $1870$ K \cite{Waal} which is reproduced by our Cleri-Rosato
crosspotential within the range of $1800 \pm 100$ K.
In order to adjust $\Delta H$ in the Al-Pt potential (which is proportional
to the strength of the interaction and to the heat of alloying in the AlPt alloy) 
the preexponential parameter $\xi$ in Eq (3) is set to $\xi \approx 3.0$ \cite{Sule_SUCI}.
Adjusting $\xi$ in Eq (3) one can tune the depth of the crosspotential well which is proportional
to $\Delta H$ \cite{Sule_SUCI,Sule_NIMB04}.

  We find, however, that the interdiffusive features of Pt in Al does not depend significantly on heat of mixing built in
the potential in accordance with our earlier findings for ion-bombardment induced intermixing in Ti/Pt \cite{Sule_NIMB04}.
The injective (ballistic) mixing of Pt takes place on a broad range of 
$\Delta H$ values including purely repulsive crosspotential 
($\Delta H=0$, $\xi \approx 0$, attractive term given in Eq. (3) is cancelled).
This is because anomalous diffusion might not be driven by thermodynamic bias (athermal mixing)
\cite{Sule_PRB05}.
The real driving force of athermal (ballistic) mixing is far from being understood clearly yet \cite{Sule_PRB05,Sule_NIMB04}.

  Depositing Pt on Al(111) using Ercolesi-Adams (EA) Al-Al potential based on the embedded atomic method \cite{EAM,EA} we get also injection
when the cross pair-potential is obtained by the Johnson's scheme \cite{Johnson} which reads as
\be
  V_{AB}(r)= \frac{1}{2}\biggm[\frac{\rho_B}{\rho_A} V_A(r)+\frac{\rho_A}{\rho_B} V_B(r)\biggm].
\ee
Hence the heteronuclear pair potential is given as the function of the elemental potentials.
To avoid singularities, cutoff must be imposed for the density functions $\rho_A$ and $\rho_B$ equal or
greater than the cutoff distance for the $V_A(r)$ and $V_B(r)$ potentials.
The cross-embedding function is given as the elemental average of $F[\rho_{i}]$ functions.
The EA potential has been fitted to {\em ab initio} atomic forces of many different
configurations using the force matching method \cite{EA}.
Recently, it has been shown, that the EA potential correctly describes
thermal expansion behavior \cite{EA} and anharmonic effects at Al surfaces 
\cite{Forsblom,premelting}.
Hence the transferability of the EAM potentials to nonequilibrium conditions of Al
seems to be sufficient.

\section{The atomic size-mismatch concept}

  Since we study the dependence of atomic dynamics on ASM we briefly
discuss our approach to ASM.
 To account for ASM the lattice mismatch (LM) concept has widely been used in the
literature which is given as the ratio of the lattice constants of the film and
substrate constituents \cite{Schukin,Tersoff,Sule_submitted,Okamoto,Jalali,Dalmas,Li,Sheng,Qi,Park}.
In the present paper we use another quantity to describe size-mismatch in diffusion
couples.
The motivation for this is that LM seems not to explain interdiffusion
and surface alloying in a number of binary systems \cite{Buchanan}.
Also, the observed asymmetry of intermixing in various metal film/substrate
couples can not be understood within the LM concept \cite{Sule_submitted,Buchanan}.
In the LM concept it is assumed that intermixing and many other properties
of binary systems (e.g. growth modes) are primarily determined by the ratio of atomic sizes (lattice
constants) of the pure elemental phases.
Instead we introduce the quantity atomic size mismatch (ASM) which is the
ratio of the first neighbor distance $r_0$ of the deposited impurity in 
its alloy phase with the substrate ($r_0^{im,s}$, AlPt alloy in this case)
to the $r_0$ of the substrate in its pure elemental phase ($r_0^s$).
If no alloy exist between the constituents than 
$r_0^{im,s}$ is simply the average of the elemental $r_0$ values.
Hence in the ASM concept the ratio $\delta_{ASM}=r_0^{im,s} / r_0^s$
depends on the cross-interaction between the impurity and substrate atoms.
It is reasonable because the impurity atom when deposited on the substrate interacts directly
with the substrate atoms hence mainly the heteronuclear interaction determines ASM at
low impurity concentration
and not its pure elemental homonuclear $r_0^{im}$ value.
With increasing film thickness and film coverage the lattice constant of the film
will be more and more important ingredient of the overall lattice misfit.
In other words within this picture of ASM the atomic volume (size) is proportional
to the local $r_0$ value which is strongly interaction and environment dependent. 
The same impurity can be described by different $r_0$ value in its pure and
in an intermixed environment.
In a strict sense the average $r_0=(r_0^{im}+r_0^s)/2$ is the correct description of
the heteronuclear first neighbor distance in the 1:1 alloy phases.
In nonstochiometric alloys or in inhomogeneous phases (such as intermixed
nonequilibrium phases) $r_0^{im,s}$ fluctuates around the elemental average and
locally $r_0^{im,s}$ can reach different values depending on the local stochiometricity
of the system. Nevertheless, the average $r_0$ seems to be a reasonable approximation
for interfacial mixing \cite{Sule_SUCI,Sule_submitted}.

 In Pt/Al the magnitude of intermixing is not sensitive to the choice of $r_0$ for the deposition of Pt on Al
within the range of $[2.8;3.0]$ $\hbox{\AA}$, hence the elemental average of $r_0^{im,s} \approx 2.85$
 $\hbox{\AA}$ is a reasonable choice.
 We prefer to use in most of the presented results the CR potential
because the first neighbor distance $r_0$ can be tuned in the heteronuclear potential in this case which is proportional to ASM.
Using EAM potentials such as given in Eq. (6) no effect of $r_0$ can be studied.

\section{Results and Discussion}

 In the rest of the article we present results which seem to be rather surprising and unexpected.
Since we find no reason on the methodological side to discard them (the employed simulation
approach and the interaction potentials should be adequate for the problem) we feel it important
to share these atomistic results with a wider community.

  The simulation of vapor deposition leads to an unexpected result. The 
deposited atoms, independently of the energy of deposition, intermix with 
ultrafast atomic exchange entering the top Al(111) layer even at 0 K within a ps if initialized $< 4-5$ $\hbox{\AA}$ far from the surface. 
Also, the particle not only enter but becomes and interstitial atom between the two upper layers.
The acceleration of the
deposited particles have been found starting at $\sim 5$ $\hbox{\AA}$ distance from the surface.
The injection of Pt goes together with the ejection of one Al atom to the surface.
The deposition of 1 ML of Pt leads to the formation of an adlayer rich in Al
in agreement with the experimental findings \cite{Buchanan,Barna}.
The ultrafast interdiffusion of Pt to Al is rather surprising, and similar
superdiffusion during soft landing of atoms on solid surfaces has been reported 
only for Pt on Cu(111) \cite{Sprague}.
The animation of the atomic injection can be seen in a web page \cite{web}.
The unexpected mechanism of Pt interdiffusion might not be an artifact of the CR potential.
We have tested recently the various surface features of CR Al potential \cite{Sule_SUCI} and got fairly nice results.
Hence, we do not attribute the anomalous behavior of Pt soft landing on Al(111) to the
artifact of the CR potential.
We find direct injection only in the case of certain transition metal elements around Pt in the periodic table,
such as Ir, Au. Other elements, such as Cu intermixes to Al after the deposition of
dozens of atoms. 
Moreover, we do believe that the fast reactive interdiffusion reported in Pt/Al and
in other diffusion couples 
must be due to anomalous diffusion which is overlooked in the literature.
We get also a rapidly increasing intermixing length during the simulation of
vapor deposition \cite{web}, although still we are far from the experimental
value of $\sim 50$ $\hbox{\AA}$ \cite{Buchanan} (results are not shown in this paper).
We deposit few MLs of Pt which is insufficient to reach this value.
In this article, however, it is not our intention to reproduce the experimental
intermixing length. We rather focus on the understanding of the elementary atomic migration step
of interdiffusion.
We will examine in the rest of the paper 
the details of the mechanism 
of the ballistic
interdiffusion of Pt.

  Although we study the mechanism of an atomic jump, we discuss
the details of an atomic transport step as a first step of 
the reactive diffusion to the bulk reported in many papers \cite{Buchanan,Barna}.
A number of such single atomic jumps leads to a diffusional process
through the atomic layers of the substrate leading to the
large intermixing length reported for Al \cite{Buchanan}.
We find the nonlinear (parabolic) scaling of $\langle R^2 \rangle$ as a function of time
during interdiffusion which suggests that the atomic inter-layer transport is anomalous
\cite{Anomaldiff}.
An atomic transport is anomalous if the exponent $\alpha > 1$ or $\alpha < 1$ in the expression
$\langle R^2 \rangle \propto t^{\alpha}$, where, $t$ is the time variable \cite{Anomaldiff}.

\begin{figure}[hbtp]
\begin{center}
\includegraphics*[height=5.5cm,width=6cm,angle=0.]{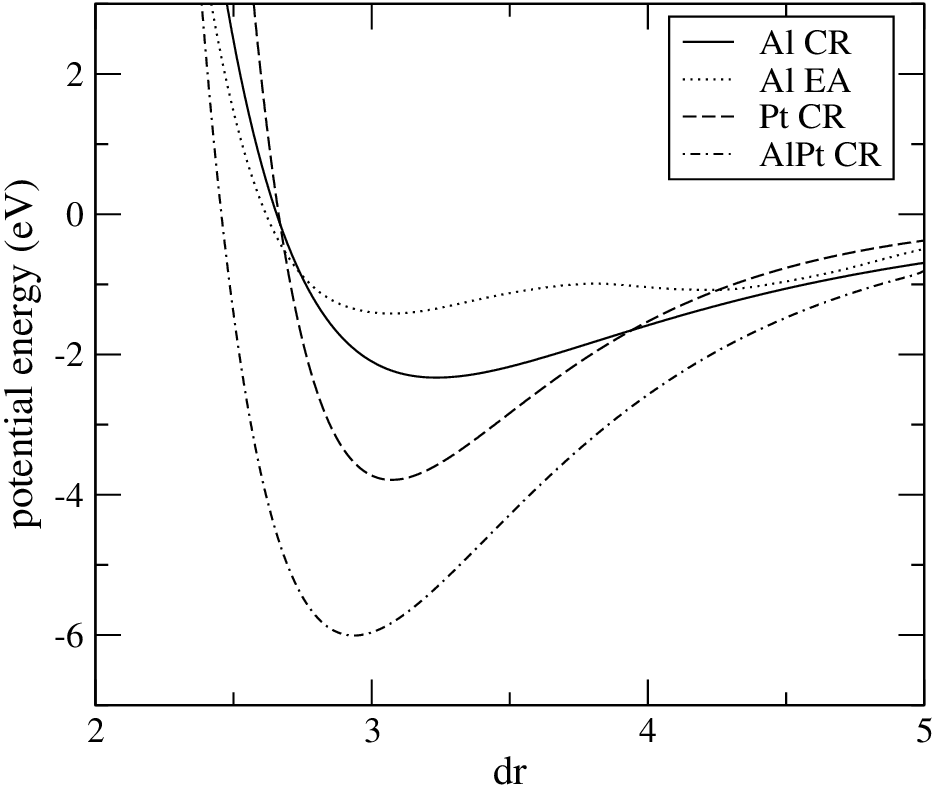}
\includegraphics*[height=5.5cm,width=6cm,angle=0.]{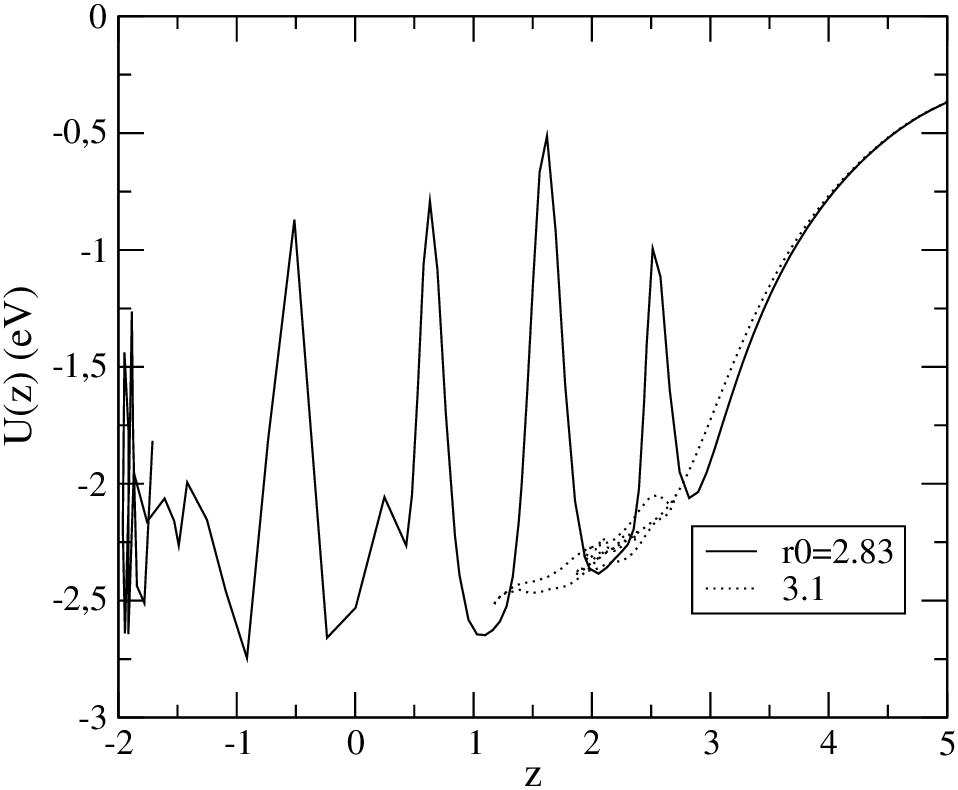}
\caption[]{
{\em Upper panel}:
The profiles of the potential energy as a function of the internuclear
separation ($r_{ij}$, $\hbox{\AA}$, for Al, Pt and for AlPt using the CR ($\xi=3.0$ in Eq. (3)) and 
Ercolesi-Adams EAM (EA)
potentials (Al only).
{\em Lower panel}: 
The oscillatory potential energy profile of Pt (straight line) as a function of the distance ($z$) of the Pt atoms from the surface
of Al(111) (the position of the surface is at $0$ $\hbox{\AA}$, $r_0 \approx 2.83$ $\hbox{\AA}$).
The potential (binding) energy of Pt with $r_0 \approx 3.1$ $\hbox{\AA}$
is given with a dotted line.
}
\label{fig1}
\end{center}
\end{figure}

 We also vary the first neighbor distance $r_0$ in the heteronuclear
potential, in order to point out the sensitivity of interdiffusion to ASM.
%$r_0$ in the crosspotential clearly reflects the size-mismatch of the pair-interaction.
The $r_0$ of the heteronuclear interaction is constructed as the average of the 
elemental values \cite{Sule_PRB05,geometric}.
As we have already pointed out,
the size-mismatch of the binary system is proportional to the ratio of
$\delta_{ASM}=r_0^{AlPt}/r_0^{Al}$.
%Hence, the atomic size difference between the elemental values is reflected
%in the average $r_0$.
In this particular case the Pt atom is the smaller particle ($r_0 \approx 2.8$)
and
$r_0 \approx 2.9$ in Al. Moreover Al has a highly anharmonic homonuclear interaction
potential (see upper Fig ~\ref{fig1}).
Due to the anharmonic behavior of Al there is a tendency for increased Al-Al
distances ($r_0^{Al} \ge 2.9$) on the surface (see upper Fig ~\ref{fig1}).
%  Indeed, we find that with decreasing $r_0$, $\langle R^2 \rangle$ more and more
%strongly diverges from linear scaling.
 If we set $r_0 \approx 3.1$ $\hbox{\AA}$, no injection is found (This is because $\delta_{ASM} \approx 1$).
%In these events $\langle R^2 \rangle$ peaks
%somewhat sooner than in
%the case of
%injective
%------------------------------------------------------
%------------------------------------------------------
%events.
Ballistic atomic injection can be suppressed by tuning ASM.
We also test deposition in other size-mismatched systems, such as Ni/Au.
Injection occurs in this system, when unphysical values are used ($r_0 \le 2.0$ $\hbox{\AA}$).

 In order to understand the details of the 
injective anomalous atomic transport
  the shape of the deposit-surface interaction energy potential is calculated.
In lower Fig ~\ref{fig1} we show the binding energy (potential energy) of the Pt atom as a function of the
distance from the surface.
  An oscillatory impurity-substrate interaction potential is found. The amplitude of the oscillation
is large (few eVs). The oscillation of the potential is caused by the
transient out-of-plane movements of surface Al atoms.
We see also the reflection of the captured Pt atom from the second Al monolayer (at $z \approx -2$ $\hbox{\AA}$).
Oscillatory interaction profiles have already been reported on the surface of metals
between
adatoms \cite{oscill}, however, no reports have been found for interdiffusion.

  When the Pt approaches the surface one or two Al atoms are released from the surface
which, however, immediately return back to the top layer.
We show the enlarged animation of the active region during injection, which
can be find at a web page \cite{web}.
It is more or less clear from this animation that the release of few
Al atoms to the surface induces the injection of Pt. 
Finally, the injection of the Pt goes together with the ejection of
an Al adatom and the Pt atom is captured below the top layer.
Prior to the injection we can see a specific mechanism. Two or three Al atoms
are released ballisticaly towards the approaching Pt atom and return back
to the surface in less then 0.1 ps (this Al atom "receives" the host atom).
The maximum vertical amplitude of the transient Al atoms can reach $\sim 2$ $\hbox{\AA}$.  
This ultrafast out-of-plane vibration of the surface Al atoms do not lead
to injection yet.
Typical 
This strongly anharmonic behavior of the Al surface is shown
in the upper panel of Fig. ~\ref{fig2} where the trajectories of the transient vertical jump of
few Al atoms to the surface (adlayer) can be seen.
The sharp potential energy wells in lower panel of Fig ~\ref{fig1} also occur due to the transient out-of-plane processes.

 Haftel {\em et al.} reported an unconventional exchange mechanism for
Au/Ag(111) diffusion process involving multiple substrate atoms which
catalyze interdiffusion \cite{Haftel2}.
They used, however, a reversed annealing simulation technique which speeds up atomic
jumps. Nevertheles, they could conclude that the activation energy of this specific
mechanism is very low.
Moreover a collective double exchange process has been found which can be activated 
ballisticaly by an incoming deposited atom.
These features are very similar to that found for the Pt/Al system.
However, we repeated the simulation for Au/Ag(111) (without reversed annealing)
at nearly $0$ K using the same simulation conditions outlined for Pt/Al, and found no injection of Au to Ag(111).
Also, at elevated temperatures (simulations have been carried out up to $300$ K) the system does not show up any interdiffusive behaviors up to few picoseconds
(not even on a longer time scale).
However, if we set $r_0 < 2.8$ $\hbox{\AA}$ (if we take the elemental average, $r_0 = 2.9$ $\hbox{\AA}$), Au does intermix to the Ag(111) topmost layer
within $2$ ps.

  Hence few diffusion couples are only available until now, including
Pt/Cu \cite{Sprague} and the
Pt/Al(111) (also Ir,Au/Al(111)) systems presented in this work which show apparent anomalous 
intermixing behavior.
Few other systems, such as the Au/Ag(111) couple could be the subject of
a classification of anomalously interdiffusing systems.
Under the effect of forced conditions, such as the low-energy ion-sputtering
of the Pt/Ti interface we also find super-interdiffusion (ballistic jumps) of Pt in the Ti substrate and also
the nonlinear scaling of $\langle R^2 \rangle$ has been obtained from simulations \cite{Sule_submitted}.
The anomalous nature of interdiffusion in Pt/Ti has also been confirmed by Auger electron spectroscopy
depth profiling as a well-defined long range tail in the concentration profile of Pt \cite{Sule_submitted}.
This finding provides further evidence for the anomalous mass transport behavior of Pt in various media.
In most of the diffusion couples with no interfacial anisotropy (atomic mass and 
size isotropy) no anomalous features have been found \cite{Sule_submitted}
and vanishingly small intermixing length have been measured and calculated \cite{Sule_PRB05}.
However, we do believe
that anomalous intermixing and impurity caused disordering of the surface of various solids are more
widespread in nature than previously thought.
In order to understand the details of the anomalous atomic transport we
study the motion of atoms following their trajectories during the
deposition events.

%We have repetaed the simulation of Au deposition on Ag(111) and found no

\subsection{Atomic trajectories in the upper layers}

\begin{figure}[hbtp]
\begin{center}
\includegraphics*[height=5.cm,width=6.cm,angle=0.]{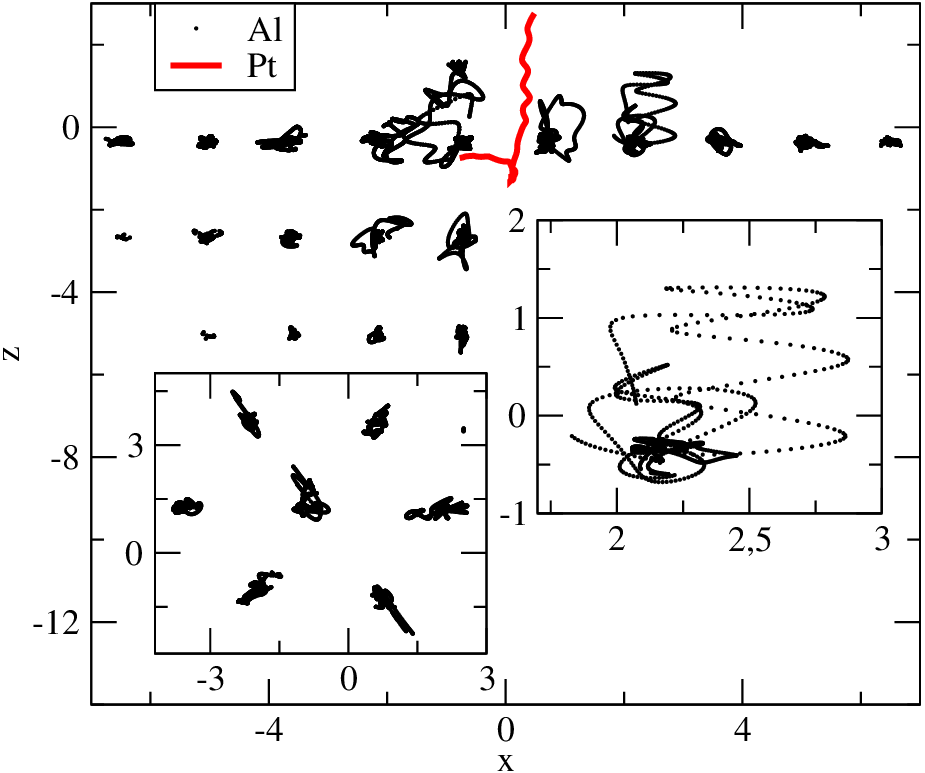}
\includegraphics*[height=5.cm,width=6.cm,angle=0.]{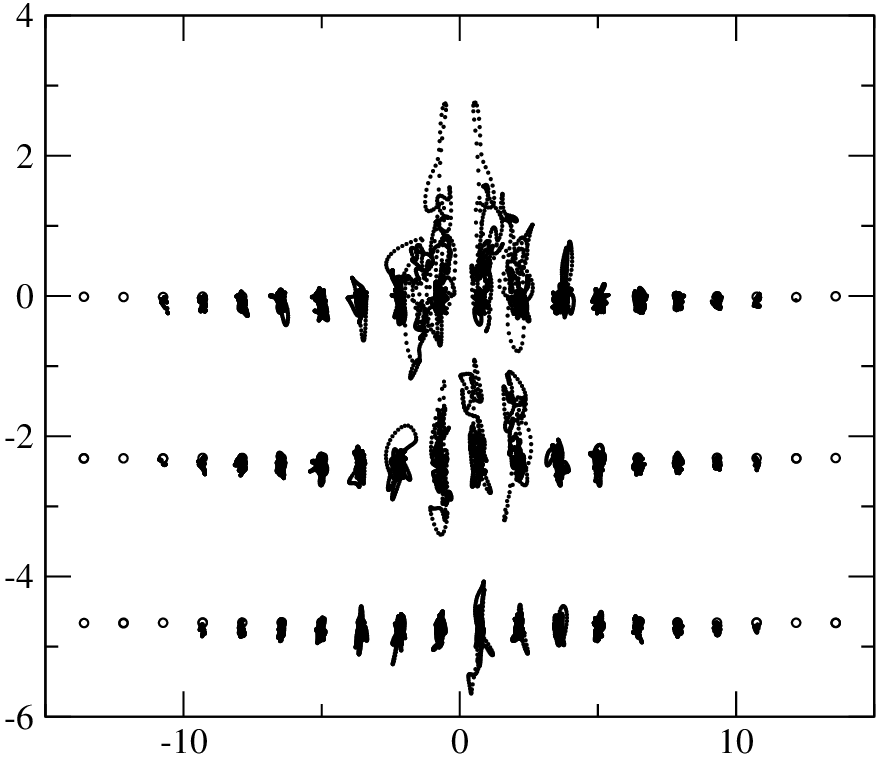}
\includegraphics*[height=5.cm,width=6.cm,angle=0.]{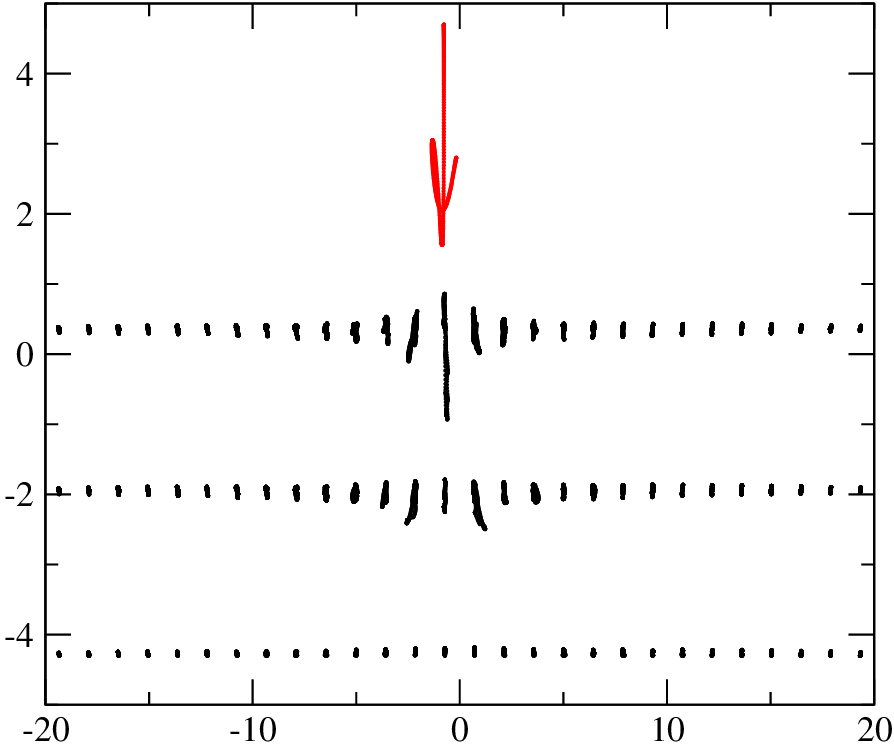}
\caption[]{
The crossectional view of typical trajectories of surface Al atoms induced
by the deposition of a Pt atom at 0 K (the $xz$ cut of the simulation cell, the scale
is in $\hbox{\AA}$ in the axes).
The positions of the Al atoms are collected up to 1 ps.
The  first neighbor distance $r_0=2.8$ $\hbox{\AA}$.
{\em Inset on the right:} The enlargement of a typical pattern of atomic dynamics of a surface
Al atom as a crossectional view.
{\em Inset on the left:} Top view (seen from the (111) surface) of lateral atomic trajectories in the central region of the
surface in the top layer.
The displaced Al atoms are arranged in a hexagonal lattice.
{\em Middle Figure:} Chaotic trajectories obtained under the same conditions
given above except that the first neighbor distance $r_0$ is set to $2.5$ $\hbox{\AA}$.
Initial atomic positions are also shown with larger filled circles.
{\em Lower Figure:} Atomic trajectories obtained for $r_0=3.1$ $\hbox{\AA}$
}
\label{fig2}
\end{center}
\end{figure}

  In order to understand more deeply the mechanism of atomic injection of Pt,
we plot the atomic positions of a crossectional slab cut in the middle of the simulation cell (with a slab thickness of $15$ $\hbox{\AA}$) for the top layer atoms during vapor deposition 
of Pt with the average $r_0 \approx 2.8$ $\hbox{\AA}$ of the elemental values (upper Fig ~\ref{fig2}) 
at $0$ K.
%------------------------------------------------------
%------------------------------------------------------
%The unexpected mechanims of Pt interdiffusion might not be an artifact of the CR potential.
The out-of-plane movement of Al atoms
is rather strong during injection, while no such movements (or with only much smaller amplitude) can be seen
when injection is suppressed by setting $r_0=3.1$ $\hbox{\AA}$ (the lower panel of Fig ~\ref{fig2}).
We conclude that the presence of the impurity atom is necessary for disordering the surface layer locally.

  The positions of the energetic atoms are also shown at $\sim 0$ K simulation in Fig ~\ref{fig2} 
We can see 
a chaotic dynamics and in certain cases turbulent or circulating atomic motions around the
equilibrium lattice sites.
In certain cases the circulating atomic trajectories could also be characterized by vortices (see the enlarged image of the pattern in the right inset of the upper panel of Fig ~\ref{fig3}).
The turbulent movement of atoms leads to large lateral
and out-of-plane amplitudes of atomic vibrations.

 We see no such state of surface atoms in the pure Al simulation cell nor in the
cases when $r_0$ is set to above a critical value ($r_c > 3.1$ $\hbox{\AA}$).
The presence of a Pt atom is necessary for the emergence of the surface disordered matter.
Moreover, the onset of the
disordered atomic motions seem to be a necessary condition of injection of Pt atoms
(this will be shown in subsection C).
The injection of Pt requires the collective motion of few vibrating surface atoms.
The assistance of at least two transient substrate adatoms with large vertical amplitudes
seems to be a necessary condition for the injective mechanism.

  In the upper panel of the left inset Fig ~\ref{fig2} the top view of the atomic trajectories of a central vibrating atom is shown surrounded
by $6$ other atoms which suggests that
the injection of Pt requires the collective motion of few vibrating surface Al atoms
with a hexagonal symmetry.
Interestingly the lateral motion of these atoms is anisotropic: the atoms
follow an elongated nearly one-dimensional trajectories.
Another atoms surrounding the $7$ "active" Al atoms are much less affected
by the impurity atom, and the 2nd neighbors are nearly unperturbed lattice
atoms on the surface.
Hence, we can say, that when injection occurs atomic vibrations should be coupled 
with each other in order to open up a channel (an empty surface site) for Pt.

  Tuning $r_0$ we also observe the onset of chaotic atomic trajectories. 
In the middle panel of Fig. ~\ref{fig2} it can be seen that the atomic motions become extremely chaotic when $r_0 \le 2.5$ 
$\hbox{\AA}$.
Further decrease in $r_0$ give rise to the ejection of
a pair of Al atoms to the vacuum.
This kind of an approach raises the question whether at least a partly chaos assisted mechanism
could be responsible for the anomalous atomic transport.
The analysis of the Lyapunov exponent (that is the measure of the chaotic processes) of the system \cite{Frenkel} could give the answer for this
question. Preliminary results indeed indicates that the presence of the
surface impurity induces atomistic chaos on the surface of Al(111) \cite{SuleTel}. 
These results will be published elsewhere.

  The local disordered surface state persists until nearly less than a ps.
Also, we find the ejection of an Al atom towards the vacuum (becomes a sputtered atom) when the EA EAM potential is used.
Hence, there are similiraties and differences between the CR and EAM potentials.
Using the CR potential we find no sputtered Al atoms and usually a single Al atom
is ejected to the surface while with the EA Al and Johnson's cross EAM potential
2-4 atoms or more are ejected to the surface and one of them is released to the
vacuum.
Using either the simple average of the elemental potentials as a heteronuclear potential or the crosspotential obtained
by the Johnson's scheme \cite{Johnson} we also find injection with the EA EAM potential.
Hence, the injective superdiffusion of Pt is not the specific artificial
feature of the CR potential but is rather a generic feature of the EAM-model.
The validity of the predicted superdiffusive features should be checked
by more sophisticated {\em ab initio} MD approach. This could be an apparent
work to be done in the future.
Although, {\em ab initio} calculations can be carried out for a substrate with the number of atoms up to $\sim 100$ 
with present day supercomputers, therefore the obtained results could also be treated
with care.

\subsection{Acceleration towards the surface}

 The deposition of impurity atoms over the substrate surface leads to the
local acceleration of the particle and to the impact to the surface with few eVs kinetic
energy even if initiated by zero velocity.
   Hence the phenomenon of acceleration of the deposited impurity atom precedes the injection
  which has already been a well-known
phenomenon \cite{Fichthorn,Sprague,Villarba,Lee}.
%------------------------------------------------------
\begin{figure}[hbtp]
\begin{center}
\includegraphics*[height=6cm,width=6.5cm,angle=0.]{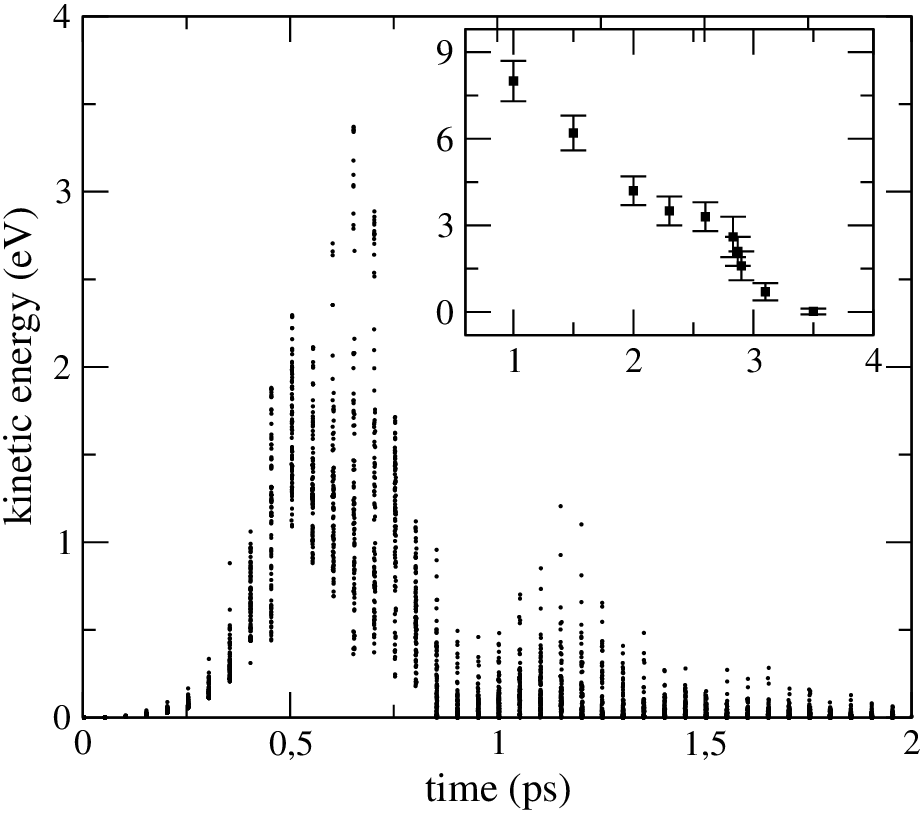}
\caption[]{
%{\em Upper panel}:
The kinetic energy of the accelerating deposited particle as a function of the time (ps)
using the Cleri-Rosato tight-binding potential ($r_0=2.83$ $\hbox{\AA}$ in the crosspotenti
al.
%using the Ercolessi-Adams EAM and Cleri-Rosato potentials
%(inset Figure) for attractive and repulsive potentials.
The particle is initialized $4$ $\hbox{\AA}$ above the free (111)
surface of the Al with nearly zero velocity.
The plotted points have been collected during
few dozens of events  with randomly varied impact points within
the $(x,y)=+1;-1$ ($\hbox{\AA}$) region of the surface.
{\em Inset:} The peak kinetic energy of the accelerating deposited particle Pt
before the impact to the surface (eV) as a function of the first neighbor
distance ($r_0$) in the heteronucelar potential.
The error bars denote standard deviations obtained for few simulation events
at each of the $r_0$ points.
}
\label{fig3}
\end{center}
\end{figure}
%------------------------------------------------------
Recently, Lee {\em et al.} demonstrated that the low-energy deposition of
Ni on Al(001) leads to serious acceleration of the Ni atoms striking the
Al(001) surface with $3-4$ eV kinetic energy \cite{Lee}.
In another study Wang {\em et al.} found the acceleration of Au particles
over the Ag(110) surface. Moreover they also find the intermixing of the Au atoms
with the Ag substrate in $11$ \% of the events in the first $6$ ps of the simulations \cite{Fichthorn}.
In the rest of the events the Au atoms remain above the surface.
In certain cases
acceleration of deposited particles leads to impact mixing (injection, e.g. in Pt/Cu)
and it has been explained by the large negative heat of solution of this
system \cite{Sprague}.
Although this conclusion is based simply on the comparison of few diffusion couples
which have different heat of solution.
 However, we demonstrate in this article that not heat of mixing ($\Delta H_m$, or heat of solution) is
the decisive factor of anomalous mixing.

 We find that even if an artificially repulsive heteronuclear potential is used ($\xi=0$ in Eq. (3)) impact mixing 
does occur for Pt/Al or Pt/Cu couples.
Also, local acceleration does occur with a weakly attractive interaction when e.g. a noble gas Ar atom
is deposited on Al (in this case no mixing occurs) with a peak
kinetic energy of $\sim 2$ eV.

  The acceleration of the deposited particle can be followed in Fig ~\ref{fig3} when
the initial velocity is zero.
The kinetic energy of the accelerating particle is given in eV as a function of time
using the CR potential (obtained for few tens of events) the peak value of $\sim 3$ eV is found when a Pt atoms
is deposited with zero initial velocity. 
The variation of the strength of the heteronuclear interaction in the potential ($\Delta H_m$)
does not affect the speed of acceleration and the peak values of the kinetic energy,
e.g., if we set the preexponential $\xi=10.$ in Eq. (3),
we set in a huge value of heat of mixing
in the Al-Pt potential (and a very strong internuclear attraction with a very deep potential energy well in the potential), however,
we find that the acceleration of Pt is not affected by the deeper potential
well in the crosspotential.
Also, if we set in smaller values, or even $\xi=0$ could be set in (purely repulsive potential, only the Born-Mayer term is nonzero given in Eq. (4)), no
change in acceleration is observed.
Hence we conclude that the speed of acceleration is not affected by the strength
of the Al-Pt potential and the generally accepted opinion, that the particle
acceleration above solid surfaces is due to the large heat of solution of the 
corresponding alloy system, is not supported \cite{Sprague}.  
In other words we find that not the thermodynamic bias, built in the cross-interaction
(via the depth of the potential well) drives the acceleration of the particle during deposition.

  The variation of $r_0$ in the crosspotential does influence, however, the speed of
acceleration. We find a correlation between $r_0$ and the peak kinetic energy of
the particle (see inset upper Fig. ~\ref{fig3}).
Although there is some scatter in the data, however, the trend is evident.
Adjusting the size-mismatch the kinetic energy of the particle impact can be tuned
by varying the relative positions of the minima of the potential wells (the ASM).
At the average $r_0$ of the elemental values ($\sim 2.83$ $\hbox{\AA}$) the abrupt change of the kinetic energy
can be seen which reports us that the system is increasingly sensitive to the variation of $r_0$ at the
physically realistic values of ASM.
Decreasing $r_0$ we find the amplification of surface instability and the
enhancement of out-of-plane vibrations.
The emergence of the injective mixing mechanism could be due to the ASM
induced surface lattice instability (local impurity induced heating up the surface).

  No acceleration of a standing particle is observed when the atom is initialized
from above $6$ $\hbox{\AA}$ the surface. We find the same decay distance of
the surface long range forces when the cutoff distance of the crosspotential
is increased above this value.

  Depositing Al atom on Al(111) acceleration also occurs (self-atomic acceleration).
The most of the deposition events with various host-substrate couples
lead to accelerative deposition regardless to the chemical nature of the
interacting couples. 
%------------------------------------------------------
\begin{figure}[hbtp]
\begin{center}
\includegraphics*[height=6cm,width=6.5cm,angle=0.]{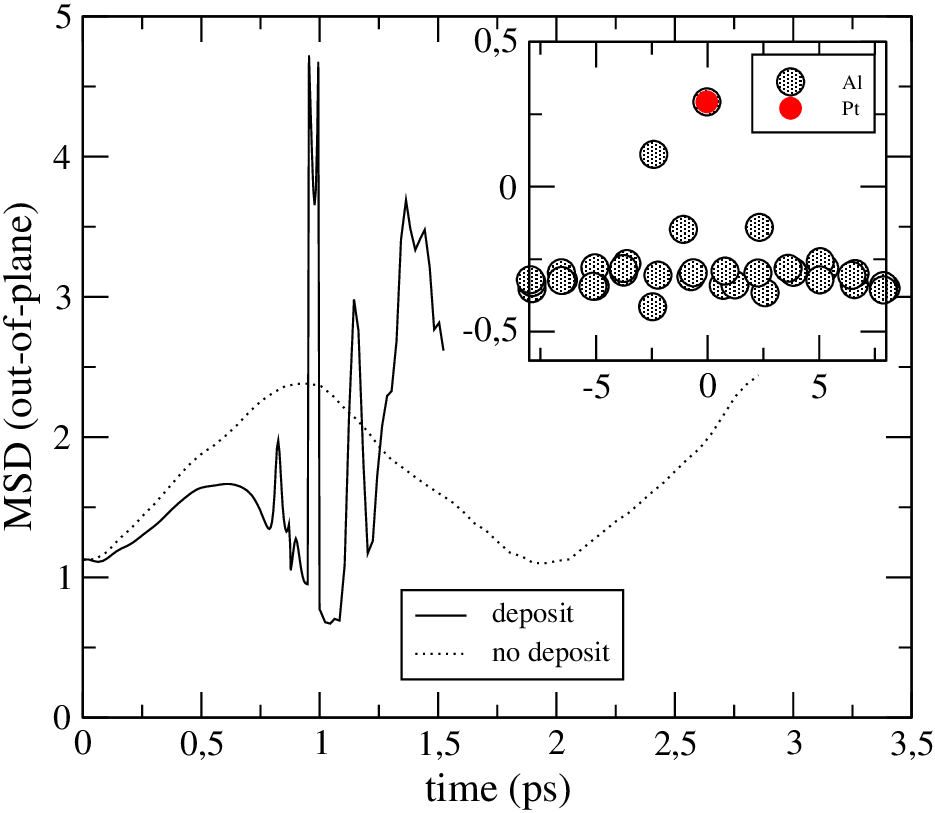}
\caption[]{
The mean square of out-of-plane atomic displacements (MSD, $\langle R^2 \rangle_z$, $\hbox
{\AA}^2$) of few Al atoms
in the topmost layer
in the injection zone ($5 \times 5$ $\hbox{\AA}^2$) as a function
of simulation time (ps) during deposition and without deposition (surface waving)
using the Cleri-Rosato potential.
The deposited atom is initialized $4.6$ $\hbox{\AA}$ above the surface.
{\em Inset:} The crossectional view of the uppermost layer at $t=0.95$ ps with
the approaching Pt atom (together with the transient Al atoms).
}
\label{fig4}
\end{center}
\end{figure}
%------------------------------------------------------

  The amplification of the out-of-plane vibration of Al atoms during deposition
gives rise to the oscillation of the electronic density $\rho$ of few surface atoms and to a long-range
density tail above the surface.
The long range forces (which initialize and speed up the acceleration) could be induced by the long range tail of the embedding
function as a function of the density $\rho$
in Eq. (2).
Due to the out-of-plane instability of Al the electron density decays slowly
above the surface and which gives rise to slowly damping $F(\rho)$.
This could lead to the enhancement of the impurity-surface interaction.

\subsection{The enhancement of surface disordering}

\begin{figure}[hbtp]
\begin{center}
\includegraphics*[height=6cm,width=6.5cm,angle=0.]{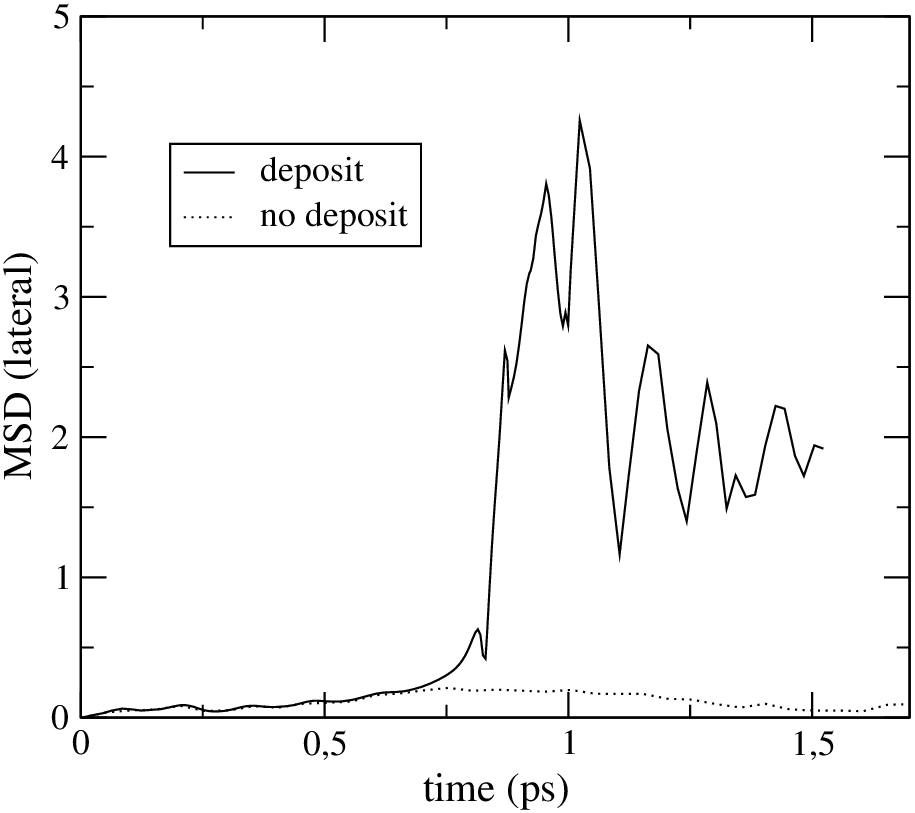}
\caption[]{
The mean square of lateral atomic displacements (MSD$_{xy}$,$\hbox{\AA}^2$) of few Al atoms
in the topmost layer
in the injection zone ($5 \times 5$ $\hbox{\AA}^2$) as a function
of simulation time (ps) during deposition and without deposition
using the Cleri-Rosato potential.
The deposited atom is initialized $4.6$ $\hbox{\AA}$ above the surface.
%{\em Inset:} The deposited particle (111) surface distance ($\hbox{\AA}$) as a function of the
%time (ps).
}
\label{fig5}
\end{center}
\end{figure}

 In order to point out the correlation between the vibrational amplitude of the
surface atoms and the acceleration of the deposited Pt atom we plot in Fig ~\ref{fig4} 
the mean square of atomic out-of-plane displacements $\langle R^2 \rangle_z$ of few surface Al atoms in the middle of the
simulation cell on the surface (using a $5 \times 5$ $\hbox{\AA}^2$ area) 
as a function of the time (ps).
In Fig ~\ref{fig5} we plot the lateral $\langle R^2 \rangle_{xy}$ in order to demonstrate the
enhancement of lateral movements of the surface Al atoms in the "active" region when
the injection takes place.
%In middle panel of Fig  we can see that the abrupt decrease of the deposit-surface distance
%as a function of the simulation time
%and which
%coincides with the abrupt increase in the vertical atomic displacements of the Al atoms.
This, together with Fig ~\ref{fig5} confirms that the abrupt activation of a surface disordering mechanism is
necessary for the injection of the Pt atom.
In Fig ~\ref{fig4} the abrupt increase of $\langle R^2 \rangle_z$ at $\sim 1$ ps has been found which
indicates the appearance of transient Al adatoms on the surface (see inset Fig ~\ref{fig4}).
This transient process coincides with the injection of Pt and catalyzes intermixing
in a similar way as proposed by Haftel {\em et al.} \cite{Haftel2}.

%------------------------------------------------------
%------------------------------------------------------

 The injection of the impurity particle starts with the acceleration of the Pt atom.
%The close proximity of the impurity atom to the surface ($d_{PtAl} \le 5$ $\hbox{\AA}$)
%induces the out-of-plane instability of the surface atoms starting first with 
%the waving of the surface which gives momentum to the Pt atom even if it was in rest.
In turn the approaching particle induces the amplification of the out-of-plane vibration
of few Al atoms which further accelerates the impurity particle towards the surface.
Finally, at a critical proximity to the surface ($d_{PtAl} \le 3$ $\hbox{\AA}$)
strong lateral components to the surface atomic displacements sets in (the appearance of the disordered surface state)
together with the out-of-plane components
and which leads to the injection of the particle.
Without the surface instability no injection occurs, the particle becomes an adatom.
The concerted motion of few surface atoms opens up a channel for injection. 

  The amplification of the out-of-plane vibration of Al atoms during deposition
gives rise to the oscillation of the electronic density $\rho$ of few surface atoms and to a long-range
density tail above the surface.
The long range forces (which initialize and speed up the acceleration) arise from the long range tail of the embedding
function as a function of the density $\rho$
in Eq. (2).
Due to the out-of-plane instability of Al the electron density decays slowly
above the surface and which gives rise to slowly damping $F(\rho)$.
This could lead to the enhancement of the impurity-surface interaction.

   Concerning the generalization of the superdiffusive features of
intermixing in Pt/Al, we can say that injection occurs in many other transition
metal/Al systems. However, in certain cases (e.g. Cu/Al) we find that the injection
of the deposited particles appears after the deposition of dozens of atoms.
In these cases, the slight decrease of $r_0$ leads also to ultrafast injection.
We see similar features in the case of the Ni/Au system \cite{web}.
As already mentioned above, injection also akes place in the Pt/Cu system in a very
similar way as in Pt/Al.
It might be the case that Pt shows superdiffusive features even in more media (substrate).
It is also interesting, what happens if we invert the systems, e.g. the
deposition of Al on Pt leads to the lack of interdiffusion during the time scale
we can reach.
However, if we set $r_0 = 2.1$ $\hbox{\AA}$, the deposited Al atom can be forced to
interdiffuse to the Pt phase \cite{web}.
Above this value no injection occurs.
It must also be noted that in general we find a transition
from superdiffusion towards normal diffusion, e.g. when the first neighbor distance
is tuned.
Finally we should emphasize that the observed anomalous atomic transport processes
and phenomena
(injection with superdiffusion, surface assisted acceleration, oscillatory deposit-surface
interaction, disordered surface matter) might not be the artifact of the employed
interaction potentials nor the PARCAS code.
The agreement with experimental vapor sputter deposition results \cite{Buchanan,Barna}
as well as the accurate parameterization of the employed Al many body potentials
\cite{Sule_SUCI,CR,EAM,EA} and the straightforward usage of them for the anharmonic effects of Al \cite{premelting,Villarba}
provide solid basis for sharing these results with the wider community.

%  We propose the following possible experiment for detecting directly the acceleration and attractive
%long range forces of the Al(111) surface.
%Using an atomic force microscope (AFM) tip with an embedded nanoscale Pt tip, an AFM scan on the Al(111) surface
%could result in important informations on the amplified interaction between the Pt
%nanoparticle and the Al surface. 
%Preliminary results indicate
%that the deposition of nanoscale Pt clusters on Al leads also to anomalous intermixing.
%The enhanced vertical movement of the tip could indicate
%the presence of effective attractive forces which force the tip for en enhanced vertical
%vibration and which results in a higher measured surface roughness
%than with another appropriate reference tip.
%The onset of weak chaos have been confirmed recently by the detection of chaotic tip 
%oscillations by AFM \cite{AFM}.
%In our case, the presence of anomalous tip-surface forces (interaction) due to oscillative
%interactions (such as shown in Fig 2.) could also lead to tip oscillations.

 An important question remains to be resolved is whether a specific state of
matter is found or the size-mismatched host-impurity interaction induced surface disordering
can be considered as a local low-temperature premelting phenomenon.
Anyhow, the impurity induced low-temperature disordering of the surface layers including few dozens of
atoms is an unexpected short lived local state of the surface and calls for further verifications.
Although its lifetime is short (less than a ps), however, this localized
spatial and temporal surface state occurs at each impact events.
This state of matter resembles a cascade event, however, its appearance
does not depend on the impact energy of the impurity particle.

\section{Conclusions}

 In this paper we studied the deposition of Pt
on Al(111). An oscillatory adsorbate-surface interaction and ultrafast injective inter-layer atomic
transport
of the deposited vapor atoms
has been found for the Pt/Al couple using two different types of atomic interaction potentials.
The soft landing of the deposit induces the disordering of the local region of the
surface Al atoms. 
The disordered state persists up to $\sim$ $0.5$ ps which covers the
transient out-of-plane vibration of few surface Al atoms.
The anomalous inter-layer transport of the impurity atom to the substrate is assisted by
the transient out-of-plane motion of few surface Al atoms.
This collective exchange mechanism works even at very low temperatures slightly above $0$ K
and calls for experimental verification.
%The acceleration of the deposited particles has also been explained by
%the long range attractive force field induced by the local waving of the surface.

 Contrary to the general belief that chemical and thermodynamic forces govern particle
deposition and interdiffusion
 no effect of thermodynamic bias has been found on acceleration towards the substrate's surface and on intermixing of the deposited particle.
Instead we point out the role of atomic size-mismatch (ASM) in particle acceleration and in superdiffusive features.
In particular we find that intermixing
can be tuned by a system parameter
which can be given as
the ratio of the atomic size parameter of the cross-interaction
term to the
substrate's lattice constant (the ASM).
Moreover, the chaotic nature of atomic trajectories as well as the
lattice instability of the surface can also be tuned
by ASM.

%\section{acknowledgment}
{\scriptsize
This work is supported by the OTKA grant F037710
from the Hungarian Academy of Sciences.
We wish to thank to K. Nordlund, to P. Barna, G. \'Odor, L. Vitos, T. Kov\'acs, T. T\'el and to M. Menyh\'ard 
for helpful discussions.
The help of the NKFP project of
3A/071/2004 is also acknowledged.
The work has been performed partly under the project
HPC-EUROPA (RII3-CT-2003-506079) with the support of
the European Community using the supercomputing
facility at CINECA in Bologna.
}

\end{document}